\documentclass[a4paper,12pt]{article}
\usepackage{graphicx}

\begin{document}
\topmargin 0pt \oddsidemargin 0mm

\renewcommand{\thefootnote}{\fnsymbol{footnote}}
\begin{titlepage}
\begin{flushright}
INJE-TP-07-09\\
\end{flushright}

\vspace{5mm}
\begin{center}
{\Large \bf Instability of  agegraphic dark energy models}
\vspace{12mm}

{\large   Kyoung Yee Kim, Hyung Won Lee and Yun Soo Myung\footnote{e-mail
 address: ysmyung@inje.ac.kr}}
 \\
\vspace{10mm} {\em  Institute of Mathematical Science and School
of Computer Aided Science \\ Inje University, Gimhae 621-749,
Korea }

\end{center}

\vspace{5mm} \centerline{{\bf{Abstract}}}
 \vspace{5mm}

We investigate the agegraphic dark energy models which were
recently proposed to explain the dark energy-dominated universe.
For this purpose, we calculate their equation of states and
squared speeds of sound.  We find that the squared speed for
agegraphic dark energy is always negative. This means that the
perfect fluid for agegraphic  dark energy is classically unstable.
Furthermore, it is shown that the new agegraphic dark energy model
could describe the matter (radiation)-dominated universe in the
far past only when the parameter $n$ is chosen to be $n>n_c$,
where the critical values are determined to be
$n_c=2.6878(2.5137752)$ numerically. It seems that the new
agegraphic dark energy model is  no better than the holographic
dark energy model for the description of the dark energy-dominated
universe, even though it resolves the causality problem.

\end{titlepage}
\newpage
\renewcommand{\thefootnote}{\arabic{footnote}}
\setcounter{footnote}{0} \setcounter{page}{2}

\section{Introduction}
Observations of supernova type Ia suggest that our universe is
accelerating~\cite{SN}. Considering the ${\rm \Lambda}$CDM
model~\cite{SDSS,Wmap1}, the dark energy and cold dark matter
contribute $\Omega^{\rm ob}_{\rm \Lambda}\simeq 0.74$ and
$\Omega^{\rm ob}_{\rm CDM}\simeq 0.22$ to the critical density of
the present universe. Recently the combination of WMAP3 and
Supernova Legacy Survey data shows a significant constraint on the
equation of state (EOS) for the dark energy, $w_{\rm
ob}=-0.97^{+0.07}_{-0.09}$ in a flat universe~\cite{WMAP3,SSM}.

Although there exist a number of dark energy models~\cite{CST},
the two promising candidates are the cosmological constant  and
the quintessence scenario~\cite{UIS}. The EOS for the latter is
determined dynamically by the scalar or tachyon.

On the other hand, there exists interesting models of the dark
energy which satisfy  the holographic principle, especially the
entropy bound. One is the holographic dark energy model~\cite{LI}
and the other is the agegraphic dark energy model~\cite{CAI}. The
first is based on the energy bound $E_{\rm \Lambda} \le E_{BH} \to
L^3 \rho_{\rm \Lambda}\le m_{\rm p}^2L$~\cite{CKN,myung} with the
length scale $L$~\footnote{Here, two length scales are introduced:
the future event horizon $R_{\rm
FH}=a(t)\int_{t}^{\infty}\frac{dt'}{a(t')}$ and the particle
horizon $R_{\rm PH}=a(t)\int_{0}^{t}\frac{dt'}{a(t')}$ with the
flat Friedmann-Robertson-Walker metric $ds^2_{\rm
FRW}=-dt^2+a^2(t)d{\bf x}\cdot d{\bf x}$. } (IR cutoff) of the
universe, while the latter is based on the K\'{a}rolyh\'{a}zy
relation of $\delta t=\lambda t^{2/3}_{\rm
p}t^{1/3}$~\cite{Karo,ND,Sas} and the time-energy uncertainty of
$E_{\delta t^3}\sim t^{-1}$ in the Minkowiski spacetime, giving
$\rho_{\rm q}\sim \frac{E_{\delta t^3}}{(\delta t)^3}\sim
\frac{m_{\rm p}^2}{t^2}$~\cite{Maz}. Hence we find the vacuum
energy density $\rho_{\rm \Lambda}=3c^2m^2_{\rm p}/L^2$ as the
holographic dark energy density, whereas the energy density of
metric perturbations $\rho_{\rm q}=3n^2m^2_{\rm p}/T^2$ with the
age of the universe $T=\int^t_0 dt'$ as the  agegraphic dark
energy density. Here the undetermined parameters $c$ and $n$ are
introduced to describe the appropriate dark energy model.
 It seems that the agegraphic  dark energy model does not
suffer the causality problem of the holographic dark energy model
because the agegraphic  dark energy model do not use the future
event horizon. However, this model suffers from the contradiction
to describe the matter-dominated universe in the far past. Hence,
the new agegraphic dark energy model was introduced to resolve
this issue~\cite{WC1}.

The two important quantities for the cosmological evolution of the
universe are the EOS  $\omega=p/\rho$ and the squared speed of
sound velocity $v^2_s=dp/d\rho$. The first describes the nature of
evolution and the latter determines the stability of evolution.

In general, it is not easy to determine the EOS for a cosmological
model with the IR cutoff $L$. If one considers $L=H_0^{-1}$
together with the cold dark matter, the EOS may take the form of
$w_{\rm \Lambda}=0$~\cite{HSU}, which is just that of the cold
dark matter.  Fortunately, choosing the future event horizon for
the holographic dark energy model determines the accelerating
universe  when using   the continuity equation~\cite{HM}.

One may consider the linear perturbation of holographic dark
energy to test the stability of  a dark energy-dominated universe.
For this purpose, an important quantity is the squared speed
$v_s^2$ of sound~\cite{PR}. The sign of $v_s^2$ is crucial for
determining the stability of a background evolution. If this is
negative, it implies  a classical instability of a given
perturbation. It was shown that the Chaplygin gas (tachyon) have
the positive squared speeds of sound with $v_{\rm
C,T}^2=-\omega_{\rm C,T}$ and thus they are supposed to be stable
against  small perturbations~\cite{GKMPS,STZW}. However, the
perfect fluid of holographic dark energy with future event horizon
is classically unstable because its squared speed is always
negative~\cite{Myung}.

In this Letter, we investigate  the agegraphic dark energy models
including new agegraphic dark energy model by calculating their
EOS $\omega_{\rm q}$ and squared speed $v^2_q$. We compare these
agegraphic dark energy models with the holographic dark energy
model. Especially, we show that the parameter $n$ of  the new
agegraphic dark energy model is restricted to $n>n_c$,  in order
 to describe the matter
(radiation)-dominated universe in the far past.

\section{ Agegraphic dark energy model}
\subsection{Noninteracting case}
In this section we discuss agegraphic dark energy
model~\cite{CAI}. A flat universe composed of $\rho_{\rm q}$ and
the cold dark matter $\rho_{\rm m}$ is governed by the Friedmann
equation
\begin{equation} \label{fried}
H^2=\frac{1}{3m^2_{\rm p}}(\rho_{\rm q}+\rho_{\rm m})
\end{equation} and their continuity equations
\begin{eqnarray} \label{cont1}
\dot{\rho}_{\rm q}+3H(\rho_{\rm q}+p_{\rm q})&=&0,\\
\label{cont2}\dot{\rho}_{\rm m}+3H\rho_{\rm m}&=&0.
\end{eqnarray}
Introducing the density parameters $\Omega_{\rm i}=\rho_{\rm
i}/3m^2_{\rm p}H^2$, then one finds
\begin{equation}
\Omega_{\rm q}=\frac{n^2}{(HT)^2}
\end{equation}
which implies that the Friedmann equation (\ref{fried}) can be
rewritten as
\begin{equation}
\Omega_{\rm q}+\Omega_{\rm m}=1.
\end{equation}
Considering the age of the universe $T=\int^a_0\frac{da'}{H'a'}$,
its pressure is determined solely by Eq.(\ref{cont1}) with $x=\ln
a$ ~\cite{HM}
\begin{equation}
p_{\rm q}=-\frac{1}{3}\frac{d\rho_{\rm q}}{d x}-\rho_{\rm
q}\end{equation}
 which  provides the EOS~\cite{CAI}
 \begin{equation} \label{aeos}
\omega_{\rm q}=\frac{p_{\rm q}}{\rho_{\rm
q}}=-1+\frac{2}{3n}\sqrt{\Omega_{\rm q}}.
 \end{equation}
Here $\omega_{\rm q}$ is determined by the  evolution equation
\begin{equation} \label{aevol}
\Omega_{\rm q}^\prime=\frac{\dot{\Omega}_{\rm q}}{H}=-3\omega_{\rm
q}\Omega_{\rm q}(1-\Omega_{\rm q})
\end{equation}
where $^\prime$ and $\dot{}$ are the derivative with respect to
$x$ and cosmic time $t$, respectively.
 For
our purpose, we introduce the squared speed of agegraphic dark
energy fluid as
\begin{equation}
v^2_{\rm q}=\frac{dp_{\rm q}}{d\rho_{\rm q}}=\frac{\dot{p}_{\rm
q}}{\dot{\rho}_{\rm q}},
\end{equation}
where
\begin{equation}
\dot{p}_{\rm q}={\dot{\omega}_{\rm q}}\rho_{\rm q}+\omega_{\rm
q}\dot{\rho}_{\rm q}
\end{equation}
with
\begin{equation}
\dot{\omega}_{\rm q}=H \frac{\Omega_{\rm q}^\prime}{3n
\sqrt{\Omega_{\rm q}}}=H\omega^\prime_{\rm q}.
\end{equation}
 It leads to
\begin{equation}
v^2_{\rm q}=-\frac{\dot{\omega}_{\rm q}}{3H(1+\omega_{\rm
q})}+\omega_{\rm q}=-\frac{\Omega_{\rm q}^\prime}{9n(1+\omega_{\rm
q})\sqrt{\Omega_{\rm q}}}+\omega_{\rm q}.
\end{equation}
In the linear perturbation theory, the density perturbation is
described by
\begin{equation}
\rho(t,{\bf x})=\rho(t)+\delta\rho(t,{\bf x})
\end{equation}
with $ \rho(t)$ the background value. Then the conservation law
for the energy-momentum tensor of $\nabla_{\nu}T^{\mu\nu}=0$
yields~\cite{KimHS}
\begin{equation} \label{pcl}
\delta \ddot{\rho}=v^2 \nabla^2 \delta \rho(t,{\bf x}),
\end{equation}
where $T^0~_0=-\rho(t)-\delta\rho(t,{\bf x})$ and $v^2=dp/d\rho$.
For $v^2_{\rm q}>0$, Eq. (\ref{pcl}) becomes a regular wave
equation whose solution is given by $\delta \rho_{\rm
q}=\delta\rho_{\rm 0 q} e^{-i\omega t+i {\bf k}\cdot {\bf x}}$.
Hence the positive squared speed (real value of speed) shows a
regular propagating mode for a density perturbation. For $v^2_{\rm
q}<0$, the perturbation becomes an irregular wave equation whose
solution is given by $\delta \rho_{\rm q}=\delta\rho_{\rm 0 q}
e^{\omega t+i {\bf k}\cdot {\bf x}}$. Hence the negative squared
speed (imaginary value of speed) shows an exponentially growing
mode for a density perturbation. That is, an increasing density
perturbation induces a lowering pressure, supporting the emergence
of instability.

 In the case of agegraphic dark energy model, one finds from Fig. 1 that the squared speed is
always negative for the whole evolution $0 \le \Omega_{\rm q} \le
1$. That is,  we read off the classical instability of $v^2_{\rm
q}<0$ for $-1 \le \omega_{\rm q}<0$. Especially, even for
$n=0.9(<1)$, there is no  phantom phase with $\omega_{\rm q}<-1$,
in contrast to the holographic dark energy model with the future
event horizon~\cite{Myung}. Also there is no sizable difference
between $n=0.9,1.2$ and 2.0 except  slightly different loci for
$v^2_{\rm q}=-1$. Furthermore, the EOS $\omega_{q}$ is a
monotonically increasing function of $\Omega_{\rm q}$, which
implies that one could not obtain the dark energy-dominated
universe in the far future ($\Omega_{\rm q} \to 1,~ \omega_{\rm q}
\to -1)$.

\begin{figure}[t!]
\centering
   {\includegraphics{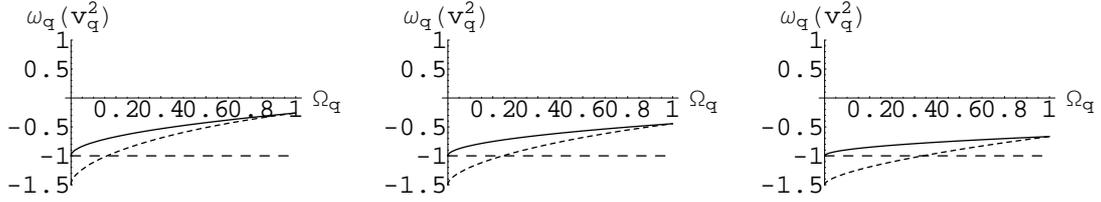}}
\caption{ Graphs for the agegraphic dark energy model. The solid
(dashed) lines denote the equation of state $\omega_{\rm q}$
(squared speed $v^2_{\rm q}$). One has three graphs for $n=0.9$,
$n=1.2$, and $n=2.0$ from the left to the right. } \label{fig1}
\end{figure}
\subsection{Interacting case}
We extend the agegraphic dark energy model to include the
interaction with the cold dark matter through the continuity
equations as \begin{equation} \dot{\rho}_{\rm q}+3H(\rho_{\rm
q}+p_{\rm q})=-Q,~~ \dot{\rho}_{\rm m}+3H\rho_{\rm m}=Q,
\end{equation}
which shows  decaying of agegraphic dark energy density into the
cold dark matter with decay rate $\Gamma=Q/\rho_{\rm q}$. In this
case, the evolution equation takes the form
\begin{equation}
\Omega_{\rm q}^\prime=\Omega_{\rm q} \Big[-3\omega_{\rm
q}(1-\Omega_{\rm q})-\frac{Q}{3m_{\rm p}^2H^3}\Big].
\end{equation}
Here we obtain the native equation of state~\cite{WGA,inter}
\begin{equation}
\omega_{\rm q}=-1+\frac{2}{3n}\sqrt{\Omega_{\rm
q}}-\frac{Q}{3H\rho_{\rm q}}.
 \end{equation}
The squared speed is given by
 \begin{equation}
v^2_{\rm q}=\frac{\omega^{\prime}_{\rm q}}{-3(1+\omega^{\rm
eff}_{\rm q})}+\omega_{\rm q},
\end{equation}
where
\begin{equation}
\omega^{\prime}_{\rm q}=\frac{\Omega^\prime_{\rm
q}}{3n\sqrt{\Omega_{\rm q}}}-\Big(\frac{Q}{3H \rho_{\rm
q}}\Big)^\prime~{\rm and}~\omega^{\rm eff}_{\rm q}=\omega_{\rm
q}+\frac{Q}{3H\rho_{\rm q}}.
\end{equation}
 If the decay rate is chosen to be $\Gamma=3 \alpha H$ with $\alpha\ge0$,
one finds a particular interaction. The evolution of the native
EOS and the squared speed are shown in Fig. 2. It seems that all
equation of states include the phantom phase with  $\omega_{\rm
q}<-1$ as well as the negative squared speed for whole range of $0
\le \Omega_{\rm q} \le 1$.
\begin{figure}[t!]
   {\includegraphics{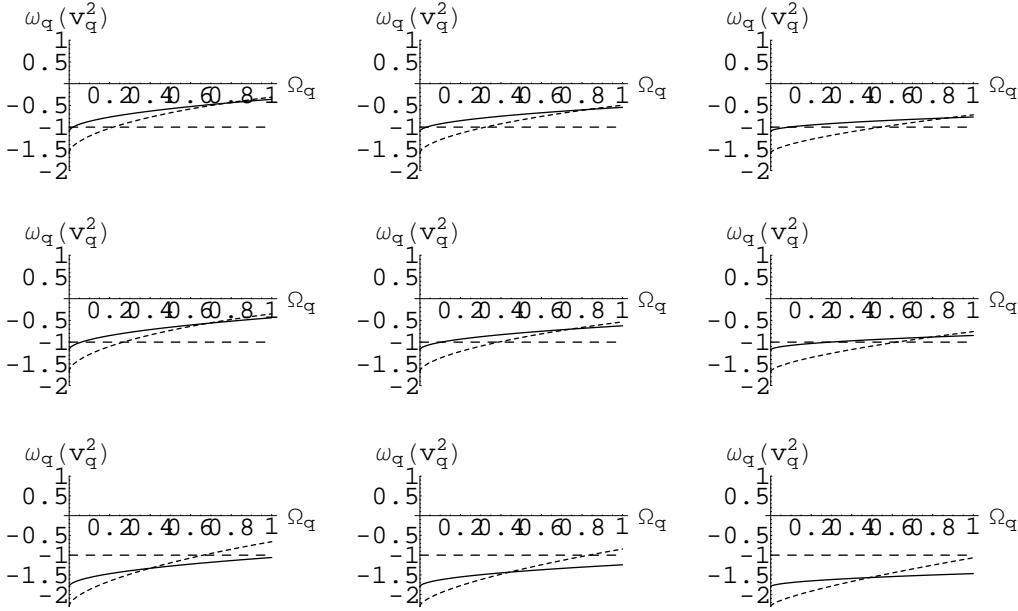}}
\caption{Nine graphs for the agegraphic dark energy. The solid
(dashed) lines denote the native  equation of state $\omega_{\rm
q}$ (squared speed $v^2_{\rm q}$). From the left to the right, one
has the graphs for $n=0.9$, $n=1.2$, and $n=2$. From  top to
bottom, one has graphs for different $\alpha=0.1,0.18$ and
$\alpha=0.8$. There is no significant change between them. }
\label{fig2}
\end{figure}
However, it is unfair to say that the interaction can induce
 the phantom phase naturally~\cite{Parv}.  In the case of interaction,
we need to introduce the effective equation of state~\cite{KLM} to
take into account the correct EOS as
\begin{equation}
\omega^{\rm eff}_{\rm q}=-1+\frac{2}{3n}\sqrt{\Omega_{\rm q}}.
 \end{equation}
In this case, the squared speed is given by
 \begin{equation}
v^2_{\rm q}=\frac{(\omega^{\rm eff}_{\rm q})
^\prime}{-3(1+\omega^{\rm eff}_{\rm q})}+\omega^{\rm eff}_{\rm q},
\end{equation}
where
\begin{equation}
(\omega^{\rm eff}_{\rm q})^\prime=\frac{\Omega^\prime_{\rm
q}}{3n\sqrt{\Omega_{\rm q}}}.
\end{equation}
\begin{figure}[t!]
{\includegraphics{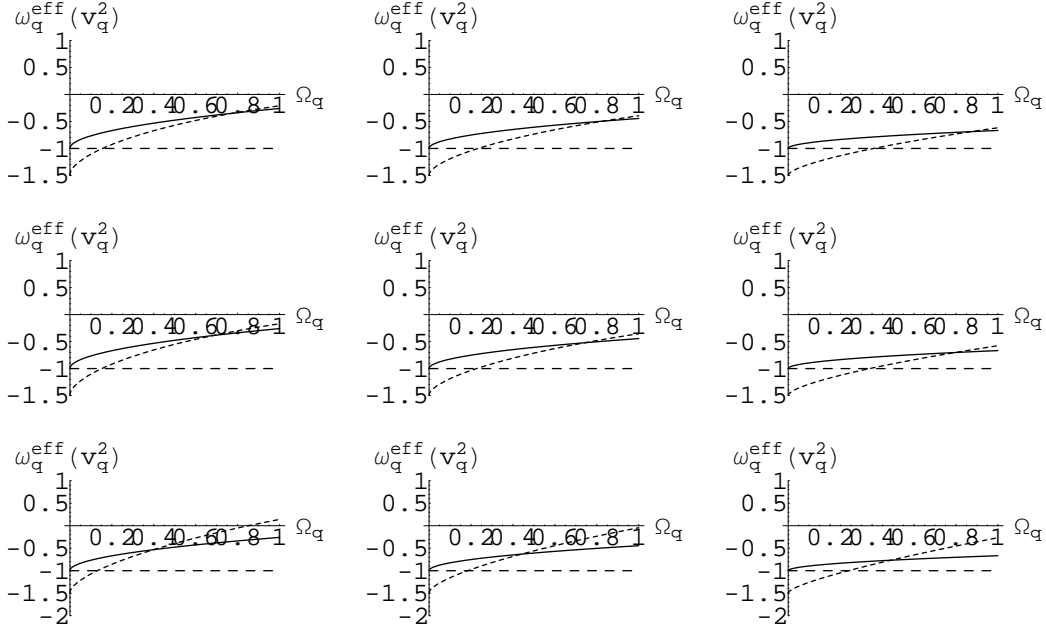}} \caption{Nine graphs for the
agegraphic dark energy. The solid (dashed) lines denote the
effective  equation of state $\omega^{\rm eff}_{\rm q}$ (squared
speed $v^2_{\rm q}$). From the left to the right, one has the
graphs for $n=0.9$, $n=1.2$, and $n=2$. From  top to bottom, one
has graphs for different $\alpha=0.1,0.18$ and $\alpha=0.8$. There
is no significant change between them.} \label{fig3}
\end{figure}
The evolution of the effective EOS and the squared speed are shown
in Fig. 3. All equation of states  are  the non-phantom phase of
$\omega_{\rm q}^{\rm eff} \ge -1$. We find that negative squared
speeds appear  for whole range of $0 \le \Omega_{\rm q} \le 1$
except $n=0.9$ and $\alpha=0.8$. Hence the agegraphic dark energy
model persists in having
 the negative squared speed,  even for
including the interaction with the cold dark matter.

\subsection{Far Past behavior of  $\Omega_{\rm q}\to 0$}
The agegraphic dark energy model has a drawback to describe the
matter-dominated universe in the far past of $a\to 0$ and
$\Omega_{\rm q} \to 0 $.  From  Eqs.(\ref{aevol}) and
(\ref{aeos}), we have $\Omega_{\rm q} \propto a^3$ and
$\omega_{\rm q}\to -1$. However, in the matter-dominated epoch, we
have $a \propto t^{2/3}$ and thus $T^2 \propto a^3$. This means
that $\rho_{\rm q} \propto a^{-3}$ and $\omega_{\rm q} \propto 0$.
On the other hand, considering the Friedmann equation
(\ref{fried}) with $\rho_{\rm m} \propto a^{-3}$, we have $H^2
\propto a^{-3}$. This
 implies that $\Omega_{\rm q} \propto {\rm const}$, which
 contradicts to $\Omega_{\rm q} \propto a^3$ in the far past.
This  inconsistency arises from the handicap of the agegraphic
dark energy model. The issue is that $\omega_{\rm q} \simeq 0$ is
not predicted  by the EOS in Eq.(\ref{aeos}). This issue is
similar to the holographic dark energy model when choosing $L_{\rm
\Lambda}=1/H_0$~\cite{HSU}, where  it may be resolved if
introducing the interaction. Here, we have to use the effective
EOS  given by $\omega^{\rm eff}_{\rm m}= -\frac{\Omega_{\rm
q}}{\Omega_{\rm m}}\frac{Q}{3H \rho_{\rm q}}$, instead of the
native EOS $\omega_{\rm m}$. In this case, choosing an appropriate
interaction $Q$ may lead to $\omega^{\rm eff}_{\rm m}\to
0$~\cite{BS}.

However, Wei and Cai have introduced the new agegraphic dark
energy model to resolve the contradiction in the far past of
agegraphic dark energy model~\cite{WC1}.

\section{New agegraphic dark energy model}
In this section, we attempt to find the EOS and squared speed for
the new agegraphic dark energy model~\cite{WC1}. This model is
different from the agegraphic dark energy model by using the
conformal time $\eta$ instead of $T$ \begin{equation}
\eta=\int^t_0\frac{dt'}{a'}=\int^a_0\frac{da'}{(a')^2H'}.
\end{equation}
Then the corresponding energy density and its density parameter
are given by
\begin{equation}
\rho_{\rm n}=\frac{3n^2m^2_{\rm p}}{\eta^2},~~\Omega_{\rm
n}=\frac{n^2}{(H \eta)^2}.
\end{equation}
\subsection{Noninteracting case}

 The pressure is determined solely by Eq.(\ref{cont1})
\begin{equation}
p_{\rm n}=-\frac{1}{3}\frac{d\rho_{\rm n}}{d x}-\rho_{\rm
n}\end{equation}
 which  provides the EOS~\cite{nadem}
 \begin{equation} \label{neos}
\omega_{\rm n}=\frac{p_{\rm n}}{\rho_{\rm n}}=-1+\frac{2
}{3na}\sqrt{\Omega_{\rm n}}=\frac{p_{\rm n}}{\rho_{\rm
n}}=-1+\frac{2 e^{-x}}{3n}\sqrt{\Omega_{\rm n}}
 \end{equation}
 with $a=e^x$.
 $\omega_{\rm n}$ is
determined by the  evolution equation
\begin{equation} \label{nequn}
\Omega_{\rm n}^\prime=-3\omega_{\rm n}\Omega_{\rm
n}(1-\Omega_{\rm n}).
\end{equation}
 Then, we introduce the squared speed of new agegraphic dark
energy fluid as
\begin{equation}
v^2_{\rm n}=-\frac{\dot{\omega}_{\rm n}}{3H(1+\omega_{\rm
n})}+\omega_{\rm n}=\frac{1}{2} (1- \Omega_{\rm n}) w_{\rm n}  +
w_{\rm n}+\frac{1}{3}.
\end{equation}

Considering  Eq.(\ref{neos})  together with the condition $a \to
0~(x \to-\infty)$ of the far past, the matter-dominated universe
is recovered with $\omega_{\rm n}=-2/3$ and $\Omega_{\rm
n}=n^2a^2/4$, while the radiation-dominated universe is recovered
with $\omega_{\rm n}=-1/3$ and $\Omega_{\rm n}=n^2a^2$~\cite{WC1}.
However, this prediction comes from the EOS $\omega_{\rm n}$ only.
We remind the reader that the pictures of far past and far future
should be determined from Eq. (\ref{nequn}) which governs the
whole evolution of the new agegraphic dark energy model.

Here we consider the noninteracting case for numerical
computations because it is a tedious calculation to explore the
whole evolution of this model including the interaction, in
contrast to the agegraphic dark energy model.  In the case of new
agegraphic dark energy model, one finds from Fig. 4 that the whole
evolution depends on parameter $n$ critically. If $n$ is less than
the critical value $n_c \sim 2.6878$, then its far fast behavior
is not acceptable because of $w_{\rm n},v^2_{\rm n} \rightarrow
\infty$ and $\Omega_{\rm n} \rightarrow 0$ as $ x \rightarrow
-\infty$. On the other hand, if $n$ is greater than the critical
value, then $w_{\rm n} \rightarrow -1$ and $\Omega_{\rm n}
\rightarrow 0$, while $v^2_{\rm n}<-1$  as $ x \rightarrow
-\infty$. In this case, we expect to have $\Omega_{\rm n} \propto
a^2$.  If $n$ approaches the critical value, then $w_{\rm n}
\rightarrow -2/3$ and $\Omega_{\rm m} \rightarrow 1$ as $ x
\rightarrow -\infty$. This corresponds to the matter-dominated
universe in the far past, predicted  by Wei and Cai~\cite{WC1}.
However, in the far future we have the convergent results of
$\Omega_{\rm n} \rightarrow 1,~\omega_{\rm n} \rightarrow
-1,~v^2_{\rm n} \to -2/3$, irrespective of $n$. Here we note that
the squared speed is always negative for $n \ge n_c$. Concerning
the evolution with respect to the redshift $z$~\cite{nadem}, this
corresponds to the evolution of region between $x=0(z=0)$ and
$x=-3.045(z=20)$ using $x=-\ln(1+z)$. As is shown in Fig. 4, we
find quite different pictures for different $n$ beyond this
region. Hence the evolution using $z$ covers a part of the whole
evolution.

\begin{figure}[t!]
\centering
\scalebox{.6}
   {\includegraphics[angle=-90]{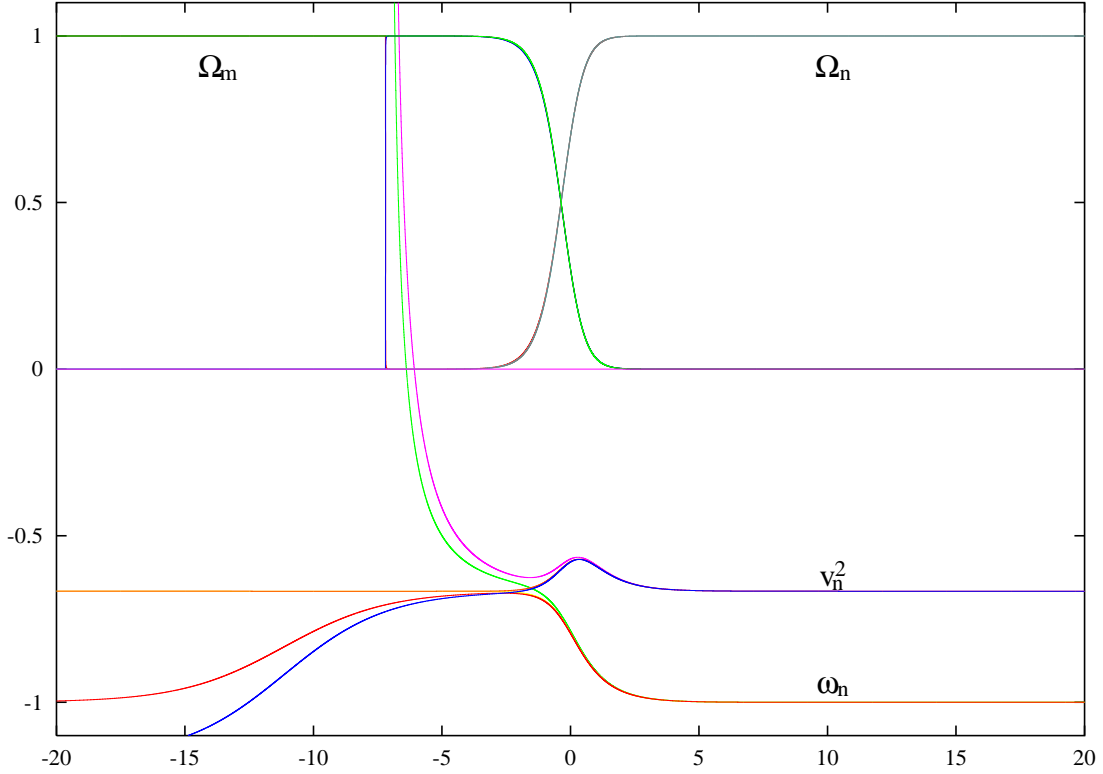}}
\caption{ Graphs for the whole evolution of the new agegraphic
dark energy. From top to down, the density parameters $\Omega_{\rm
n}(\Omega_{\rm m})$, the squared speed $v^2_{\rm n}$, and the EOS
$\omega_{\rm q}$ are depicted.  Each graph is composed of
different $n$ values, $n = 2.6, 2.6878, 2.7$. One  distinguishes
difference for different $n$ in the region of  $x<0$ and  the
upper curve in each graph corresponds to smaller  value of
$n=2.6$. } \label{fig4}
\end{figure}

In order to discuss the radiation-dominated universe in the far
past, one has to include a radiation matter which satisfies
$\dot{\rho}_{\rm r}+4H\rho_{\rm r}=0$. Then the evolution
equations for density parameters are modified as
\begin{eqnarray}
\Omega_{\rm n}' = \Omega_{\rm n} [ \Omega_{\rm r} -3 ( 1- \Omega_{\rm n}) w_{\rm n} ], \label{evolve_cdm_rad}\\
\Omega_{\rm r}' = \Omega_{\rm r} [ 3 w_{\rm n} \Omega_{\rm r} - ( 1- \Omega_{\rm r}) ].
\end{eqnarray}
Solving these equations numerically,  we find the similar graphs
in  Fig. 5 except the shift of $w_{\rm n}$ from $-2/3$ to $-1/3$
in the far past. We observe that there exist a critical value $n_c
\sim 2.5137752$, which  provides $w_{\rm n} \rightarrow -1/3$ as $x
\rightarrow -\infty$. This corresponds to the radiation-dominated
universe with $\Omega_{\rm r} \to 1$. The squared speed is
obtained as
\begin{equation}
v^2_{\rm n}=-\frac{\dot{\omega}_{\rm n}}{3H(1+\omega_{\rm
n})}+\omega_{\rm n}=-\frac{1}{6} \left [ \Omega_{\rm r} - 3 (1-
\Omega_{\rm n}) w_{\rm n} \right ]  + w_{\rm n}+\frac{1}{3}.
\end{equation}
For $n=n_c$, the squared speed  leads to $-1/3$ as $x \to
-\infty$.  The critical value of $w^{\rm c}_{\rm n}=-1/3$ for the
radiation-dominated universe can be understood as follows. To have
an asymptotic value of  $w_{\rm n} = w_{\rm n}^{\rm c}$ at $x =
-\infty(a\to0)$, we have
\begin{equation}
\Omega_{\rm n} = \frac{9 n^2}{4} (1+w_{\rm n}^{\rm c} )^2 e^{2 x} .
\end{equation}
Substituting this into Eq.(\ref{evolve_cdm_rad}), one gets
\begin{equation}
\Omega_{\rm r} -3 \left [ 1- \frac{9 n^2}{4} (1+w_{\rm n}^{\rm c})^2 e^{2x} \right ] w_{\rm n}^{\rm c} = 2 .
\end{equation}
 Taking the limit of  $x \rightarrow -\infty$, we find  $w_{\rm n}^{\rm c}$  as
\begin{equation}
w_{\rm n}^{\rm c} = \frac{\Omega_{\rm r} -2}{3}.
\end{equation}
It gives $w_{\rm n}^{\rm c}=-1/3$ for $\Omega_{\rm r} = 1$
(radiation-dominated era) and $w_{\rm n}^{\rm c}=-2/3$ for
$\Omega_{\rm r} = 0$ (matter-dominated era).

\begin{figure}[t!]
\centering
\scalebox{.6}
   {\includegraphics[angle=-90]{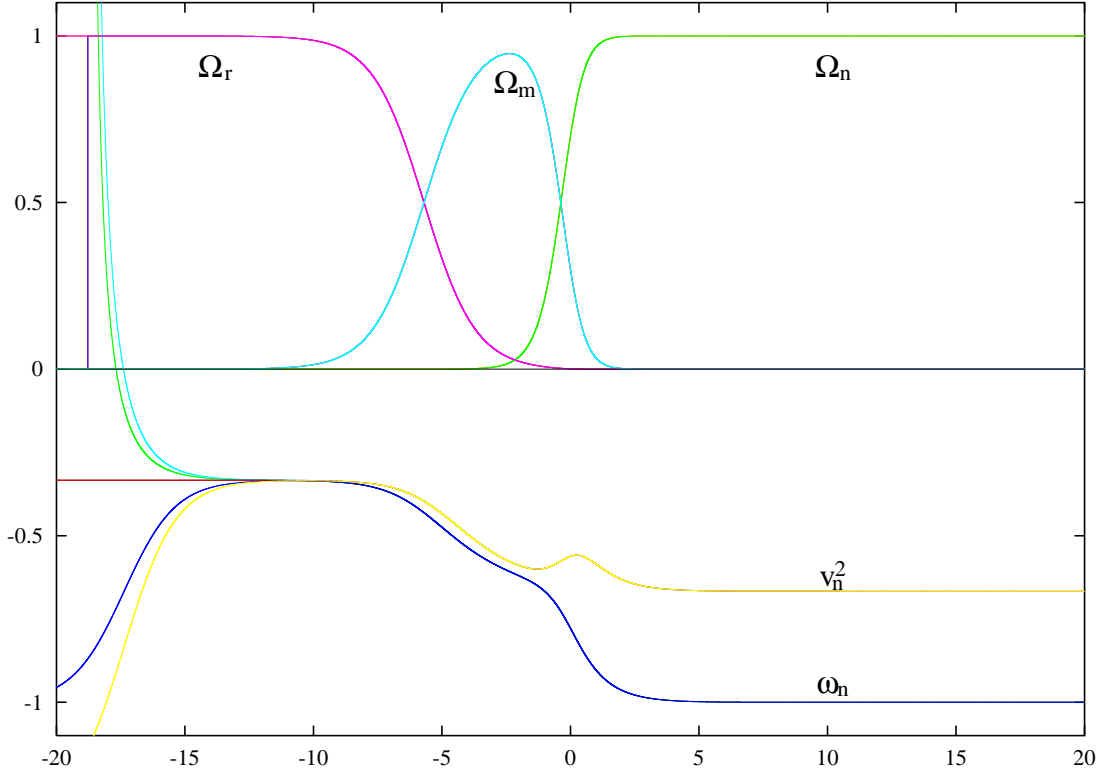}}
\caption{ Graphs for the whole evolution of the new agegraphic
dark energy. From top to down, the density parameters $\Omega_{\rm
n}(\Omega_{\rm m},\Omega_{\rm r})$, the squared speed $v^2_{\rm
n}$, and the EOS $\omega_{\rm q}$ are depicted. Each graph is
composed of  different $n$, $n = 2.513775, 2.513775196, 2.513776$.
One distinguishes difference for different $n$ in the region of
$x<0$ and the upper curve in each graph corresponds to smaller
value of  $n=2.513775$. } \label{fig5}
\end{figure}

\subsection{Interacting case}
We extend the new agegraphic dark energy model to include the
interaction with the cold dark matter through the continuity
equations as \begin{equation} \dot{\rho}_{\rm n}+3H(\rho_{\rm
n}+p_{\rm n})=-Q,~~ \dot{\rho}_{\rm m}+3H\rho_{\rm m}=Q,
\end{equation}
which shows  decaying of new agegraphic dark energy density into
the cold dark matter with decay rate $\Gamma=Q/\rho_{\rm n}$. In
this case, the evolution equation takes the form
\begin{equation}
\Omega_{\rm n}^\prime=\Omega_{\rm n}
\Big[-3(1-\Omega_{\rm n})\Big(-1+\frac{2
e^{-x}}{3n}\sqrt{\Omega_{\rm n}}\Big)-\frac{Q}{3m_{\rm
p}^2H^3}\Big].
\end{equation}
Here we obtain the native equation of state~\cite{WC1}
\begin{equation}
\omega_{\rm n}=-1+\frac{2 e^{-x}}{3n}\sqrt{\Omega_{\rm
n}}-\frac{Q}{3H\rho_{\rm n}}.
 \end{equation}
The squared speed is given by
 \begin{equation}
v^2_{\rm n}=\frac{\dot{\omega}_{\rm n}}{-3H(1+\omega^{\rm
eff}_{\rm n})}+\omega_{\rm n},
\end{equation}
where
\begin{equation}
\dot{\omega}_{\rm n}=\frac{\Omega^\prime_{\rm
n}}{3n\sqrt{\Omega_{\rm n}}}-\frac{2 e^{-x}}{3n}\sqrt{\Omega_{\rm
n}}-\Big(\frac{Q}{3H \rho_{\rm n}}\Big)^\prime~{\rm
and}~\omega^{\rm eff}_{\rm n}=\omega_{\rm n}+\frac{Q}{3H\rho_{\rm
n}}.
\end{equation}
 If the decay rate is chosen to be $\Gamma =3 \alpha H$ with $\alpha\ge0$,
one finds an interacting picture. The evolution of the native EOS
and the squared speed are similar to the case in  Fig. 4 except
that all equation of states may include the phantom phase with
$\omega_{\rm n}<-1$.

However, if we  introduce the effective equation of state
\begin{equation}
\omega^{\rm eff}_{\rm n}=-1+\frac{2 e^{-x}}{3n}\sqrt{\Omega_{\rm
n}},
 \end{equation}
 the squared speed is given by
 \begin{equation}
v^2_{\rm n}=\frac{\dot{\omega}^{\rm eff}_{\rm n}
}{-3H(1+\omega^{\rm eff}_{\rm n})}+\omega^{\rm eff}_{\rm n},
\end{equation}
where
\begin{equation}
\dot{\omega}^{\rm eff}_{\rm n}=\frac{\Omega^\prime_{\rm
n}}{3n\sqrt{\Omega_{\rm n}}}-\frac{2 e^{-x}}{3n}\sqrt{\Omega_{\rm
n}}.
\end{equation}
We expect that the evolution of the effective EOS  may be  free
from the phantom phase.  It is conjectured that negative squared
speeds may appear in the whole evolution of the universe. Hence
the new agegraphic dark energy model may persist in having  the
negative squared speed,  even for including the interaction with
the cold dark matter.

\begin{table}
\caption{Comparison between new agegraphic dark energy model
(NADEM) and  holographic dark energy model (HDEM) with the future
event horizon. Far past means $a\to0(x\to -\infty)$, while Far
future denotes $a\to\infty(x\to \infty)$.  MDU (DEDU) means
matter-dominated universe of $\Omega_{\rm m}\to 1$(dark
energy-dominated universe of $\Omega_{\rm q}\to 1$). }
\begin{tabular}{|c|c|c|c|}
  \hline
     & NADEM &  HDEM &Remark \\
  \hline
  Far past   & $\omega_{\rm n}=-\frac{2}{3}(v^2_{\rm n}=0)$ for $n=n_c$ & $\omega_{\rm \Lambda}
                  =v^2_{\rm \Lambda}=-\frac{1}{3}$ for all $c$&MDU\\
  Far future  & $\omega_{\rm n}=-1(v^2_{\rm n}=-\frac{2}{3})$ for all $n$ & $\omega_{\rm \Lambda}=v^2_{\rm \Lambda}=-1$ for $c=1$&DEDU \\
  \hline
\end{tabular}
\end{table}
\section{Discussions}

We investigate the agegraphic dark energy model and new agegraphic
dark energy model.  For this purpose, we calculate their equation
of states and squared speeds of sound . We find that the squared
speed for agegraphic dark energy is always negative. This means
that the perfect fluid for agegraphic  dark energy is classically
unstable. Furthermore, it is shown that the new agegraphic dark
energy model could describe the matter (radiation)-dominated
universe in the far past only when the parameter $n$ is chosen to
be $n>n_c$ where the critical value is determined  by
$n_c=2.6878(2.5137752)$ numerically.

As is summarized in Table I, if one considers the far past as the
matter-dominated universe, there is a slight difference in their
EOS of dark energy sector. However, this difference is meaningless
because the role of dark energy sector is not important in the far
past of matter-dominated universe. In the far future, both evolve
toward the dark energy-dominated universe. The difference is that
for new agegraphic dark energy model, one has $\omega_{\rm n}\to
-1$ for all $n$, while for holographic  dark energy model, one has
$\omega_{\rm \Lambda}\to -1$ for $c=1$ only. However, this
compensates the far past behavior: for new agegraphic dark energy
model, one has $\omega_{\rm n}\to -2/3$ for $n_c=2.6878$, while
for holographic dark energy model, one has $\omega_{\rm
\Lambda}\to -1/3$ for all $c$.

Concerning the squared speed, we find $-2/3\le v^2_{\rm n} \le 0$
for new agegraphic dark energy model with $n=n_c$, while for
holographic dark energy model with $c=1$, its range is $-1/3\le
v^2_{\rm n} \le -1$.  Hence,  for these cases, we always have the
negative squared speed, showing the instability of their fluid
models.

In conclusion,  the new agegraphic dark energy model is no better
than the holographic dark energy model for the description of the
dark energy-dominated universe, even though it resolves the
causality problem.

\section*{Acknowledgment}
  K. Kim and H. Lee were  in part supported by
KOSEF, Astrophysical Research Center for the Structure and
Evolution of the Cosmos at Sejong University. Y. Myung  was in
part supported by  the Korea Research Foundation
(KRF-2006-311-C00249) funded by the Korea Government (MOEHRD).


\begin{thebibliography}{99}
\bibitem{SN}
A. G. Riess {\it et al.}, Astron. J. {\bf 116} (1998)
1009 [astro-ph/9805201 ]; 
S. J. Perlmutter {\it et al.}, Astrophys. J. {\bf 517} (1999)
565 [astro-ph/9812133]; 
A. G. Riess  {\it et al.}, Astrophys. J. {\bf 607} (2004)
665[astro-ph/0402512]; 
P.~Astier {\it et al.},
  Astron.\ Astrophys.\  {\bf 447} (2006) 31
  [arXiv:astro-ph/0510447].

\bibitem{SDSS}
  M.~Tegmark {\it et al.}  [SDSS Collaboration],
  Phys.\ Rev.\ D {\bf 69} (2004) 103501
  [arXiv:astro-ph/0310723];

K.~Abazajian {\it et al.}  [SDSS Collaboration],
  Astron.\ J.\  {\bf 128} (2004) 502
  [arXiv:astro-ph/0403325];

  K.~Abazajian {\it et al.}  [SDSS Collaboration],
  Astron.\ J.\  {\bf 129} (2005) 1755
  [arXiv:astro-ph/0410239].

\bibitem{Wmap1} H. V. Peiris  {\it et al.}, Astrophys. J. Suppl. {\bf 148} (2003) 213
[astro-ph/0302225];
C. L. Bennett  {\it et al.}, Astrophys. J. Suppl. {\bf 148} (2003)
1[astro-ph/0302207];
D. N. Spergel  {\it et al.}, Astrophys. J. Suppl. {\bf 148} (2003)
175[astro-ph/0302209].



\bibitem{WMAP3}
  D.~N.~Spergel {\it et al.},
  arXiv:astro-ph/0603449.

\bibitem{SSM}
  U.~Seljak, A.~Slosar and P.~McDonald,
  arXiv:astro-ph/0604335.



  \bibitem{CST}
  E.~J.~Copeland, M.~Sami and S.~Tsujikawa,
  arXiv:hep-th/0603057.

\bibitem{UIS} A. Upadhye, M. Ishak, and P. J. Steinhardt, Phys. Rev. D {\bf 72} (2005)
063501[arXiv:astro-ph/0411803].

\bibitem{LI} M. Li, Phys. Lett. B {\bf 603}
(2004) 1[arXiv:hep-th/0403127].

\bibitem{CAI}
  R.~G.~Cai,
  arXiv:0707.4049 [hep-th].



\bibitem{CKN} A. Cohen, D. Kaplan, and A. Nelson, Phys. Rev. Lett.
{\bf 82} (1999) 4971[arXiv:hep-th/9803132].

\bibitem{myung}
  Y.~S.~Myung,
  Phys.\ Lett.\  B {\bf 610} (2005) 18
  [arXiv:hep-th/0412224].

\bibitem{Karo} F. K\'{a}rolyh\'{a}zy, Nuovo Cim. A {\bf 42}
(1966) 390.

\bibitem{ND}
Y.~J.~Ng and H.~Van Dam,
  Mod.\ Phys.\ Lett.\  A {\bf 9} (1994) 335.
\bibitem{Sas}
N.~Sasakura,
  Prog.\ Theor.\ Phys.\  {\bf 102} (1999) 169
  [arXiv:hep-th/9903146].



\bibitem{Maz}
  M.~Maziashvili,
  Phys.\ Lett.\  B {\bf 652} (2007) 165
  [arXiv:0705.0924 [gr-qc]].

  \bibitem{WC1}
  H.~Wei and R.~G.~Cai,
  arXiv:0708.0884 [astro-ph].

\bibitem{HSU} S. D. Hsu, Phys. Lett. B {\bf 594} (2004) 13[arXiv:hep-th/0403052].

\bibitem{HM}
  Q.~G.~Huang and M.~Li,
  JCAP {\bf 0408} (2004) 013
  [arXiv:astro-ph/0404229].

    \bibitem{PR}
     P.~J.~E.~Peebles and B.~Ratra,
       Rev.\ Mod.\ Phys.\  {\bf 75} (2003) 559
      [arXiv:astro-ph/0207347].

 \bibitem{GKMPS}
  V.~Gorini, A.~Kamenshchik, U.~Moschella, V.~Pasquier and A.~Starobinsky,
  Phys.\ Rev.\  D {\bf 72} (2005) 103518
  [arXiv:astro-ph/0504576].

   \bibitem{STZW}
  H.~Sandvik, M.~Tegmark, M.~Zaldarriaga and I.~Waga,
  Phys.\ Rev.\  D {\bf 69} (2004) 123524
  [arXiv:astro-ph/0212114].

\bibitem{Myung}
  Y.~S.~Myung,
Phys.\ Lett.\  B {\bf 652} (2007) 223
  [arXiv:0706.3757 [gr-qc]].



\bibitem{KimHS}
  H.~Kim,
  Mon.\ Not.\ Roy.\ Astron.\ Soc.\  {\bf 364} (2005) 813
  [arXiv:astro-ph/0408577].



\bibitem{WGA} B.~Wang, Y.~g.~Gong and E.~Abdalla,
  Phys.\ Lett.\  B {\bf 624} (2005) 141
  [arXiv:hep-th/0506069].

 \bibitem{inter}
  H.~Wei and R.~G.~Cai,
  arXiv:0707.4052 [hep-th];
H.~Wei and R.~G.~Cai,
  arXiv:0707.4526 [gr-qc];

\bibitem{Parv}
  D.~Pavon and W.~Zimdahl,
  Phys.\ Lett.\  B {\bf 628} (2005) 206
  [arXiv:gr-qc/0505020].




\bibitem{KLM}
H.~Kim, H.~W.~Lee and Y.~S.~Myung,
  Phys.\ Lett.\ B {\bf 632} (2006)  605
  [arXiv:gr-qc/0509040];

\bibitem{BS}
  M.~S.~Berger and H.~Shojaei,
  Phys.\ Rev.\  D {\bf 73} (2006) 083528
  [arXiv:gr-qc/0601086].





\bibitem{nadem}
  H.~Wei and R.~G.~Cai,
  arXiv:0708.1894 [astro-ph].



  \bibitem{obser}
  X.~Wu, Y.~Zhang, H.~Li, R.~G.~Cai and Z.~H.~Zhu,
  arXiv:0708.0349 [astro-ph];
  Y.~Zhang, H.~Li, X.~Wu, H.~Wei and R.~G.~Cai,
  arXiv:0708.1214 [astro-ph].


  \bibitem{Neu}
  I.~P.~Neupane,
  arXiv:0708.2910 [hep-th].

























       \end{thebibliography}
        \end{document}